\documentclass[preprint2]{aastex}

\newcommand{\simle}{\mbox{$\stackrel{<}{_{\sim}}$}}
\newcommand{\simge}{\mbox{$\stackrel{>}{_{\sim}}$}}

\slugcomment{Accepted, ApJ Jan 2005}

\shorttitle{Multi-wavelength imaging of IRC\,+10216}
\shortauthors{Tuthill et al.}

\begin{document}

\title{Multi-wavelength diffraction-limited imaging of 
              the evolved carbon star IRC\,+10216, II}

\author{P. G. Tuthill\altaffilmark{1}, 
        J. D. Monnier\altaffilmark{2}, \and 
        W. C. Danchi\altaffilmark{3}} 

\altaffiltext{1}{School of Physics, University of Sydney, NSW 2006, Australia}
\altaffiltext{2}{University of Michigan at Ann Arbor, Department of Astronomy, 
                 500 Church Street, Ann Arbor, MI 48109-1090, USA}
\altaffiltext{3}{NASA Goddard Space Flight Center, Infrared Astrophysics, 
                 Code 685, Greenbelt, MD 20771, USA }

\begin{abstract}
High angular resolution images of IRC\,+10216 taken at various
bandpasses within the near-infrared {\it h, k} \& {\it l} bands are presented.
The maps have the highest angular resolution yet recovered, and were
reconstructed from interferometric measurements obtained at the Keck~1
telescope in 1997 December and 1998 April, forming a subset of a
7-epoch monitoring program presented earlier \citep[][paper~1]{T00}.
Systematic changes with observing wavelength are found and discussed
in context of present geometrical models for the circumstellar
envelope.  With these new high-resolution, multi-wavelength data and
contemporaneous photometry, we also re-visit the hypothesis that the
bright compact Core of the nebula (component ``A'') marks the location
of the central carbon star.  We find that directly measured properties
of the Core (angular size, flux density, color temperature) are
consistent with a reddened carbon star photosphere
(line-of-sight $\tau_{2.2}=5.3$).
\end{abstract}

\keywords{stars: individual(\objectname{IRC +10216}), stars: mass loss,
techniques: interferometry}

\section{Introduction}

IRC\,+10216 (= CW~Leo) is a dusty, embedded carbon-rich long period
variable presently undergoing an episode of intense mass-loss up to
$10^{-5}$\,M$_\odot$/yr \citep{MH99}.  It is the nearest object of its
type \citep[110-135\,pc;][]{G97} and brightest in the thermal-IR, and
is generally believed to be in transition between the latest stellar
and the earliest planetary nebula phases.  This fortuitous combination
of factors has led to intensive study across the spectrum resulting in
an extensive and rich literature, making IRC\,+10216 the textbook
example for objects in its class.  Although studies of molecular lines
\citep[eg.][]{BT93} and deep imaging in scattered galactic light
\citep{MH99} reveal the outer parts of the mass-loss nebula to contain
a series of spherical shells, the innermost regions show a bipolar
structure \citep{KW94}.  As the physical processes driving the
transition from spherical to bipolar symmetry are not well understood,
there has been much interest in imaging and modeling this system at
the finest scales as a prototype addressing one of the outstanding
problems in the latest stages of the stellar life cycle.  The ability
to recover diffraction-limited images in the infrared with a
large-aperture telescope has delivered a detailed picture of the
central regions of IRC\,+10216 \citep{HB98,W98,T00,O00,W02}.  The
bright Core is shown to be highly clumpy and inhomogeneous, and with
studies now spanning some years, evolution of the material with
apparent brightening, fading, and proper motion makes for considerable
complexity in the immediate circumstellar environment.  Here we
present the highest angular resolution images yet made, spanning the
near infrared {\it h, k,} and {\it l} bands.

\section{Observations}
  
Using the techniques of aperture masking interferometry, images of
IRC\,+10216 at a range of near-IR wavelengths were obtained from data
taken at the 10\,m Keck~I telescope.  These observations formed a
somewhat distinct component of a program which also comprised the
7-epoch {\it k}~band imaging study presented in Paper~I.  This paper is
concerned with an intensive study of IRC\,+10216 made over only two
observing runs in 1997 December and 1998 April, during which images
were made in 5 separate near-infrared wavebands.  The bandpasses of
the filters used are given in Table~\ref{tbl-filters} while an
observing log showing the dates, aperture masks and filters can be
found in Table~\ref{tbl-log}.

\begin{deluxetable}{lcc}
\tablewidth{0pt}
\tablecaption{\label{tbl-filters}
Properties of Filters}
\tablehead{
\colhead{Name} & \colhead{Wavelength} & \colhead{Bandwidth} \\
 &  \colhead{($\mu$m)} & \colhead{($\mu$m)}  \\
}
\startdata
{\sf h}     & 1.657 & 0.333 \\
{\sf kcont} & 2.260 & 0.053 \\
{\sf ch4}   & 2.269 & 0.155 \\
{\sf pahcs} & 3.083 & 0.101 \\
{\sf pah}   & 3.310 & 0.063 \\
\enddata
\end{deluxetable}

\begin{deluxetable}{cllll}
\tablewidth{0pt}
\tablecaption{\label{tbl-log}
Journal of Interferometric Observations.
For consistency with Paper~1, we have preserved the nomenclature
of the epochs from the larger 7-epoch set presented earlier,
explaining why the epochs start at 2. See text and Table~1 for
descriptions of the masks and the filters.}
\tablehead{
\colhead{Epoch} & \colhead{Date} & \colhead{Mask} & 
\colhead{Filters} & \colhead{Phase\tablenotemark{1}}}
\startdata
2 & 1997 Dec 16 & Annulus & {\sf h,kcont}           & 1.23 \\
2 & 1997 Dec 16 & Golay21 & {\sf kcont,ch4,pahcs}   & 1.23 \\
2 & 1997 Dec 18 & Golay21 & {\sf kcont,pahcs,pah}   & 1.23 \\
3 & 1998 Apr 14 & Golay21 & {\sf ch4,pahcs}         & 1.41 \\
3 & 1998 Apr 15 & Annulus & {\sf h,ch4,pahcs,pah}   & 1.41 \\
\enddata
\tablenotetext{1}{Stellar Phases from \cite{MGD98}}
\end{deluxetable}

In contrast to Paper~I which was concerned with changes of the
morphology of the dust shell between the separate epochs, here we have
averaged together data taken over two separate runs.  This was done to
enhance the signal-to-noise (S/N) ratio of some of the maps, and to
compensate for the fact that there was no single observing epoch which
yielded high quality maps at all the observing wavelengths of
interest.  Although the two epochs chosen were only 4~months apart,
measurable changes in the relative locations of components in the
inner dust shell were shown to occur in Paper~I.  However, the fastest
moving component would be displaced only $\sim$8.5\,mas in this
interval (see Paper~I) -- averaging together maps with such small
shifts should produce no great bias in the results.

The Golay and Annulus aperture geometries from Table~\ref{tbl-log} are
described in detail in \cite{keckmask}, which contains a thorough
description of the experimental methods.  Although the two masks
employed did differ in performance for the various levels of source
flux and seeing conditions encountered, comparisons proved that there
were no systematic differences in the final maps produced allowing
them to be averaged together by observing wavelength alone.  As
absolute positional information is not recovered from our
closure-phase based techniques, maps to be averaged were registered
with respect to each other by maximizing the cross-correlation before
being summed.


\section{Results}

This section presents the major observational findings of this paper.
Diffraction-limited image reconstructions, visibility curves and
additional supporting data are presented and discussed.

\subsection{{\it k}~band images -- a comparison of bandwidths}

\begin{figure*}[ht]
\includegraphics[angle=90,width=18cm]{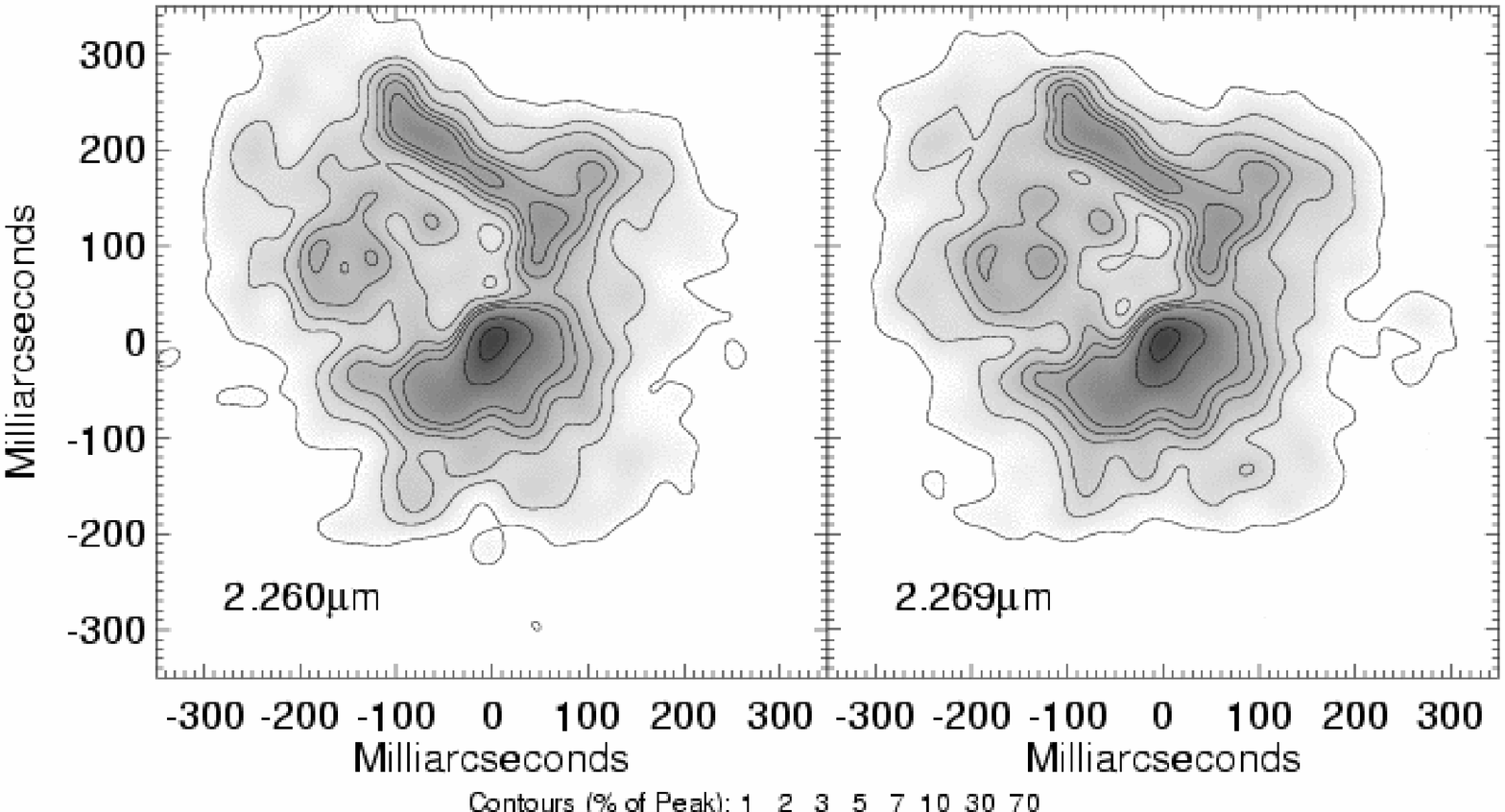}
\caption{ \label{2kmaps}
Image reconstructions of IRC\,+10216 in the near-infrared {\it k}~band taken 
in 1997 December.
{\em Left-hand panel} average of 4 maps taken with the {\sf kcont} filter;
{\em right-hand panel} average of 2 maps taken with the {\sf ch4} filter.
See Table~\ref{tbl-filters} for a description of filter properties.}
\end{figure*}

Reconstructed images of IRC\,+10216 from data taken in 1997 December
through two different filters, {\sf kcont} and {\sf ch4}, are given in
Figure~\ref{2kmaps}.  As is apparent from Table~\ref{tbl-filters}, the
{\sf kcont} and {\sf ch4} filters have similar central passbands, but 
their bandwidths differ by a factor of 3.  The two images of
Figure~\ref{2kmaps} do, however, exhibit a high degree of similarity
to each other, with no significant departures beyond those to be
expected at the lowest contours near the level of the noise.  This is
not surprising given that the emission process concerned is thermal
radiation from warm dust which exhibits a fairly featureless spectrum
across the near-infrared.  The findings from this comparison, and
other maps presented later in the paper, confirm that rapid changes in
morphology with wavelength were not seen.

Images such as those of Figure~\ref{2kmaps} also present a useful
yardstick for measuring the fidelity of the image reconstructions.
The level of agreement shown is typical for maps taken with different
aperture masks and on different nights with varying seeing conditions.
Such external consistency checks have allowed us to determine that
structures above a certain level, in this case about 3\% of the peak, 
are very well established experimentally.

\subsection{A descriptive model of IRC\,+10216}

\begin{figure}[ht]
\hspace{-1cm}
\includegraphics[angle=90,width=10cm]{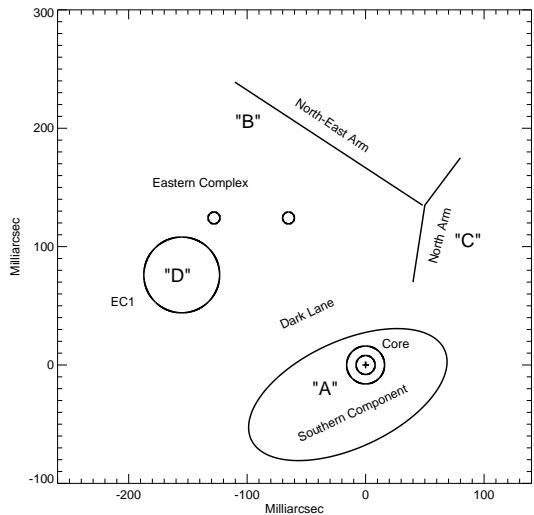}
\caption{ \label{cartoon}
Cartoon diagram of features identified in high resolution images of
IRC\,+10216 (see also Figure~2 from Paper~1).
Features discussed in the text are labelled and registered against the peak 
of the bright Core of the Southern Component (denoted by the symbol + above).
The ``A,B,C,D'' nomenclature of \citet{HB98} is also indicated.
}
\end{figure}

A qualitative model of the appearance of the dust shell around
IRC\,+10216 was given in Paper~I.
We repeat this description here, together with a `cartoon sketch'
labeling the features given in Figure~\ref{cartoon}, which also
gives the ``A,B,C,D'' nomenclature of \citet{HB98}.  
The brightest feature in the maps of
Figure~\ref{2kmaps}, appearing as a sharp spike at the map center, we
denote as the Core (``A'').  Surrounding the Core is a roughly
elliptical region of emission elongated along a position angle of
$\sim120^\circ$ which we label as the Southern Component.  The shorter
North Arm starts from a location 100\,mas NW of the Core, and extends
approximately 150\,mas to the NNW (``C'').  Perpendicular to it runs
the prominent North-East Arm (``B''), running over 200\,mas and
forming the brightest structure outside the Southern Component.
Displaced 150\,mas from the Core to the NE is a region containing
multiple weaker features, labeled as the Eastern Complex.  Within
this region, the brightest feature, which consists of a roughly
circular patch somewhat less than 100\,mas across, has been tagged as
cloud EC1 (``D'').  For further clarification of these assignments,
see Figure~2 of Paper~I.  The different nomenclature employed 
compared to other workers is necessitated by the higher angular 
resolution, so that the simple blobs reported in earlier work are
resolved into clumpy or filamentary structures here.

\subsection{Multi-wavelength near-infrared images}

\begin{figure*}[hbt!]
\includegraphics[angle=90,width=16cm]{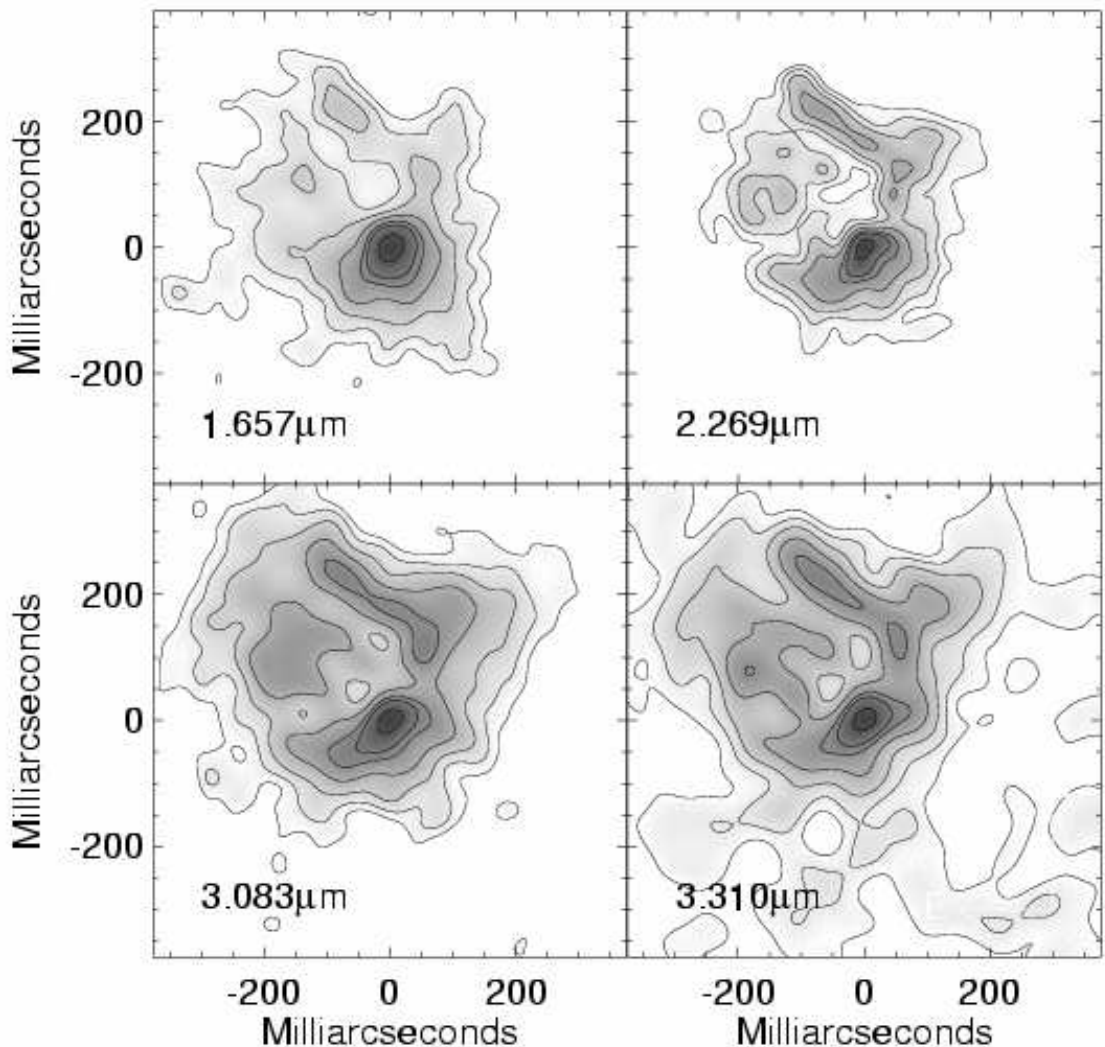}
\caption{ \label{hkllmaps}
Image reconstructions of IRC\,+10216 from data taken at four different 
wavelengths in the near-infrared over the epochs 1997 December and 1998 January.
{\em Upper left} average of 3 maps taken with the {\sf h} filter; 
{\em upper right} average of 4 maps taken with the {\sf ch4} filter;
{\em lower left} average of 8 maps taken with the {\sf pahcs} filter;
{\em lower right} average of 2 maps taken with the {\sf pah} filter.}
\end{figure*}

Maps of IRC\,+10216 spanning four near-infrared wavebands,
representing a noise-weighted average of data collected over 1997
December and 1998 April are given in Figure~\ref{hkllmaps}.
Unfortunately, the S/N and image fidelity was never as high for maps
outside the {\it k}~band.  For shorter wavelengths, a combination of lower
intrinsic flux from the star together with the greater role played by
the atmospheric seeing were probably to blame for this loss.  For the
observations at wavelengths greater than 3\,$\mu$m, the intrinsically
lower system angular resolution probably contributed to enhanced
difficulties with the mapping, although it is likely additional
unidentified factors were also at play.  For these reasons, the
contour levels in Figure~\ref{hkllmaps}, chosen to be near or above
the noise, are more conservative than for Figure~\ref{2kmaps}.
Despite this precaution, it can be seen for the map at 3.310\,$\mu$m
({\sf pah} filter) resulting from an average of only two somewhat noisy 
maps, that the lowest contours do show the effects of noise.

In examining maps for systematic differences as a function of
observing wavelength, the reader is cautioned to bear in mind that the
four images of Figure~\ref{hkllmaps} all have different angular
resolutions and noise levels.  Maps follow a similar basic shape, 
with the dominant features such as The Core, North and North-East Arms, 
and EC1 being identifiable at all wavelengths.  
However, it is apparent that changes in appearance do occur across 
the near-infrared.  

\begin{figure*}[htb!]
\includegraphics[angle=90,width=16cm]{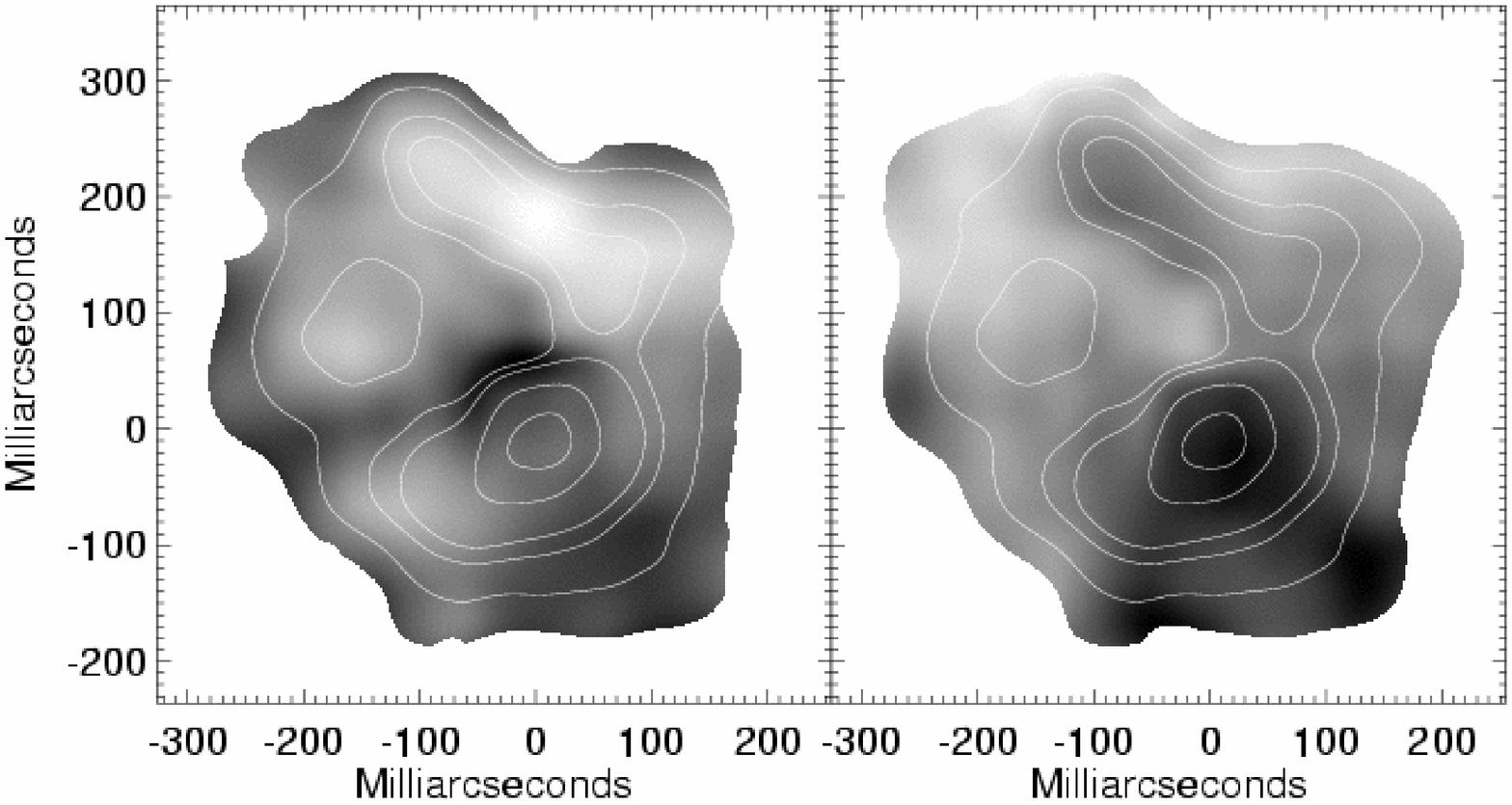}
\caption{ \label{colormaps}
Color images with the {\sf h} - {\sf ch4} filter (left panel) and 
the {\sf ch4} - {\sf pahcs} filter (right panel). 
In order to mitigate the effects of varying angular resolution for the
different filters, images were smoothed to an effective beamsize of 
70\,mas before subtraction. 
Data were also truncated at 3\% of the peak brightness giving rise to
the blank areas at the edges.
The overplotted contours show the structure from the smoothed {\sf ch4}
map, indicating the locations of the Core and dominant features.
For the {\sf h} - {\sf ch4} map, the reddest (light) areas correspond to
3.7\,mag, and the bluest (dark) 1.8\,mag.
For the {\sf ch4} - {\sf pahcs} map, the reddest areas are 3.0\,mag 
and the bluest 1.4\,mag.
However, more extreme colors may occur over small regions but will be 
diluted by the smoothing mentioned above.
}
\end{figure*}

In order to visualize these changes, Figure~\ref{colormaps} shows color
maps produced by differencing the {\sf h-k} and {\sf k-pahcs} images.
As might be expected in a source with such complexities of morphology
where opacity, temperature and scattering all play varying roles, 
interpreting such maps is not straightforward. 
Of the bright high-SNR features, the Core exhibits the bluest (hottest)
spectrum in {\sf k-pahcs}.
However, in the {\sf h-k} map, this region exhibits a flat spectrum,
neither relatively blue nor red, although the bluest feature in the
map does lie quite nearby toward the dark lane.
Even more puzzling is the behavior of the North/North-East Arm and
(to a lesser extent) EC1, which are among the reddest features in 
the {\sf h-k} map, yet appear markedly blue in {\sf k-pahcs}.
It is important to point out that among the many other physical 
properties which need to be modeled to understand these maps, there
is also a strong molecular absorption feature due to $ HCN $ and 
$ C_2H_2 $ \citep{ISO99} which coincides with the {\sf pahcs} filter
bandpass.

Further discussion of the colors and sizes of features, particularly the 
Core, is given later, however we prefer to do this based on simple 
models fit directly to the observed visibility data themselves.
This is because the reconstruction of an image from a set of visibilities 
and closure phases is a complex and nonlinear process, and can be subject 
to biases introduced by the deconvolution process.
Where possible, quantitative interpretation of the data is often best
achieved by fitting to the measurements rather than proceeding by
the intermediate step of a reconstructed image.

\subsection{Visibility data}

\begin{figure*}[ht!]
\includegraphics[width=15cm]{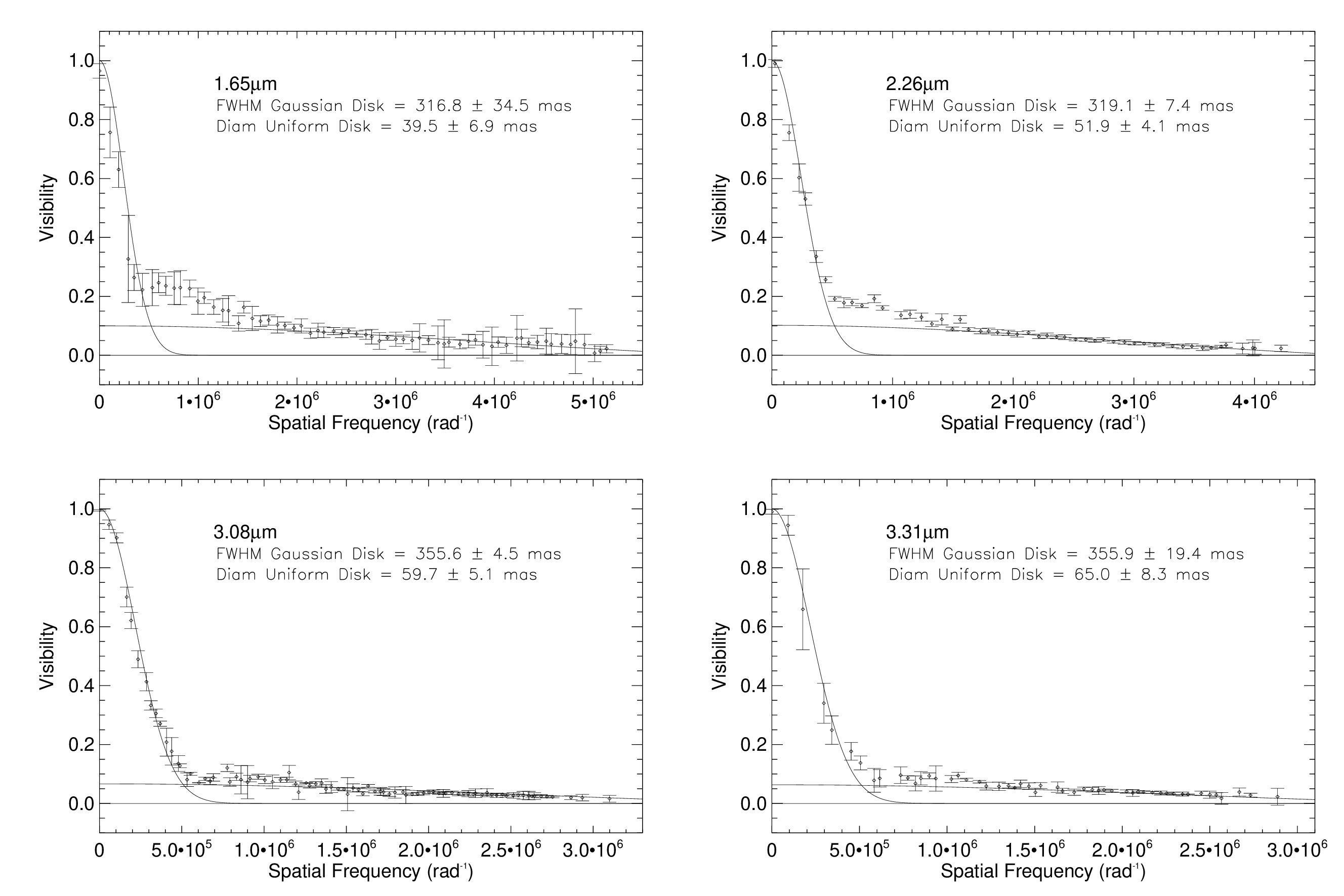}
\caption{ \label{viscurves}
Azimuthally averaged visibility data for IRC\,+10216 at four different 
wavelengths: {\em Upper left} {\sf h} filter; {\em upper right} {\sf ch4} filter;
{\em lower left} {\sf pahcs} filter; {\em lower right} {\sf pah} filter.
Two separate models (solid lines) have been fit to data over different 
spatial frequency ranges.
The rapid fall-off in visibility at low resolution (spatial frequencies
less than $4.5\times10^5~rad^{-1}$ were used for the model) has been fit with
a circular Gaussian disk model, with full-width at half-maximum (FWHM) 
given in the figure.
The gradual decline at high spatial frequencies (greater than 
$2\times10^6~rad^{-1}$ was used for the model), implying the presence
of a bright partially-resolved component, has been fit with a uniform
disk model with the diameter as given.
Errors were computed from the spread of values among separate fits to a 
number of independent datasets at each wavelength.
The best-fitting model parameters and their errors, together with 
the model visibility curves themselves, are all plotted in the figure.}
\end{figure*}

Simple quantitative brightness models were fit directly to the
calibrated visibilities.
Figure~\ref{viscurves} presents azimuthally averaged visibility
data for IRC\,+10216 at four near-infrared wavelengths.  At all
wavelengths, the visibility curves exhibit a common basic form.  A
rapid drop from high visibilities at the origin occurs over a spatial
frequency range \simle~0.5$\times10^6$\,rad$^{-1}$, denoting that most
of the flux comes from well-resolved large scale structure.  This
component has been modeled by fitting a circular Gaussian disk to the
visibility data, the results of which are shown in
Figure~\ref{viscurves}.  Intermediate spatial frequencies
(0.5$\times10^6$ -- 2.0$\times10^6$\,rad$^{-1}$) carry much of the
signal which results in the asymmetric form of the maps. Owing to this
known complexity, no attempts were made to fit to azimuthally averaged 
visibilities on these intermediate baselines.
At high angular resolutions (\simge 2.0$\times10^6$\,rad$^{-1}$) 
visibilities for all wavelengths appear to follow a simple functional 
form corresponding to a partially resolved, circularly symmetric, 
compact component.  A uniform disk model has been fit to
these data, and best fits are overplotted on the figure.  

We proceed by identifying the uniform disk component fit to the long
baselines as due to the Core. 
However, this identification requires careful argument and justification.
The idea that the Core is responsible for the highest spatial frequency
structure is to an extent confirmed by the appearance of the recovered 
images, which show this to be by far the dominant compact feature.
Furthermore, based on the physical assumption that the Core may indeed be 
the star, it is not unreasonable to expect that it is able to dominate
the high-angular-frequency spectrum over flux coming from extended dust,
which is less likely to have sharp structure on fine scales. 
In this case, mapping confirms the assumption to be reasonable, however
there may be some non-core contamination of visibilities at intermediate
baselines, a difficulty which only higher angular resolution can cure.
A second important caveat concerns the high-resolution asymmetry noted to be 
associated with the Core \citep{T00,O00}, which could cause difficulties
in fitting to azimuthally averaged data (which assumes circular symmetry).
The elongation was not as great at these epochs as it later became 
\citep{T00}, and tests were performed where data were grouped in sectors 
according to the orientation of the baseline. 
The spread of values found for these Core diameter fits was incorporated
into the error term for the final diameters given in Figure~\ref{viscurves}. 
 
\section{Discussion and Conclusions}


The purported location of the stellar photosphere within the IRC\,+10216 
nebula has been discussed in various studies starting with \citet{KW94}.
However, the complexities of the dust shell, and the absence of an 
unambiguously distinct bright compact component have caused continuing 
uncertainty over identifying which, if any, feature corresponds to the 
star itself.  This has led to fundamental uncertainties
over the structure of the dust shell, and determining the location of
the star is essential to further progress.  In Paper~I we argued that the 
Core of the southern component (``A'') consisted of emission from the stellar 
photosphere with the possible addition of some flux from hot dust on the 
inner wall of a cavity or torus, tilted into the line of sight in the south.  
While this bright Core, with an angular diameter of 50\,mas, was the prime
candidate to be the stellar photosphere, the possibility that the star
itself lay completely obscured behind the central dark dust lane was
also advanced.

\citet{O00} suggested an alternate interpretation, that the star
itself lies within the bright knot at the end of the NE arm (component
``B''). Evidence supporting this view came from color maps, from
observations of changes in shape of the Core, and particularly from
attempts to trace the point of origin of the polarization field.  An
ambitious series of papers built on this interpretation, including 
detailed radiative transfer modeling \citep{M01,MHW02} and the results 
of a further observational campaign \citep{W02}.  Detailed models
interpreting the southern component as a cavity in a bipolar structure
(``A'') with the star at the location of the North-East arm (``B'') were 
produced which gave reasonable fits to both the spectral energy 
distribution and the high resolution imaging.  
However these models were quite complex, incorporating $>$20 free 
parameters describing many different shells of varying symmetry and 
density profile.
Based on this analysis, the authors rule out the possibility that 
the star can either be at the Southern Core, or that it was completely 
hidden behind the dust \citep{M01}.

Despite this, there remain a number of simple arguments against the 
view that the star is located in the North-East arm.  At the higher
resolutions afforded by the Keck imaging, the North-East arm is seen
to be a linear structure with almost uniform brightness along a ridge
with no dominant compact knot.  Furthermore, the North-East arm was found 
to be elongating and fading with time (Paper~I), and in the most recent 
Keck observations (publication in preparation), almost absent entirely.  
Similar findings of the disappearance of the feature they previously 
identified as the star were explained by \citet{MHW02} who invoke a recent 
and dramatic increase in mass-loss rate which has had the effect of 
burying it beneath a thickening dust shell.  However our higher angular
resolution Keck observations show that the North-East arm is actually 
lengthening and separating from the Southern Component, while it 
gradually fades out.  The observed proper motion, with tentative 
acceleration detection \citep{keckmask,W02}, would not be expected
from the identifications of \citet{M01} of these two components
as a part of the circumstellar torus and the stellar photosphere.

Recent lunar occultation measurements \citep{CM01,RCL03} have been
able to shed some light on this question with extremely high angular
resolutions (0.8\,mas) attained in one-dimensional profiles.  The Core
of the southern component was found to be well fitted by a compact
Gaussian profile with FWHM 35\,mas contributing 6\% of the total flux 
in {\it k}~band.  The close agreement with the expected angular size of the
photosphere for the star, as well as the finding of a component of
similar size and position in the {\it h}~band profile, led these authors to
identify the southern Core (``A'') as the stellar disk, although not
with sufficient confidence to rule out alternative hypotheses.

By fitting only the longest baselines as shown in Figure~\ref{viscurves}, 
we were able to discriminate between the compact Core and the 
well-resolved nebula, revealing a partially-resolved component 
giving measurable visibility signal at even the longest baselines.  
For the four different bandpasses depicted (1.65, 2.26, 3.08 \& 
3.31\,$\mu$m), uniform disk fits yielded diameters of 40, 52,
60 \& 65\,mas with relative flux contributions of 10, 10, 7 \& 6\% of
the total respectively.  Our {\it k}~band uniform disk diameter of 52\,mas
is in close agreement with the lunar occultation measurements of
\citet{RCL03} whose 35\,mas FWHM Gaussian model would correspond, in
the partially resolved case here, to a uniform disk diameter of 56\,mas.  
The observed variation in angular diameter with wavelength across the 
near-IR is not necessarily strong evidence that the feature is a
dust clump.  To the contrary, it has become apparent that opacity
changes due to various molecular and atomic species in the atmospheres
of evolved stars lead to precisely such large apparent diameter
variations, at least for M-stars \citep{r_aqr,Mn02}.  

We turn now to examination of the apparent color of the compact Core. 
From the uniform disk fits above, the same relative contribution from
the Core was found in {\it h} and {\it k}~band (both 10\%), but significantly
less at the longer wavelengths (7 \& 6\% at 3.08 \& 3.31\,$\mu$m 
respectively).
This finding is in good accord with the color maps presented in 
Figure~\ref{colormaps} which found the Core to be neutral in {\sf h-k}
and blue in {\sf k-pahcs}.
If the Core is indeed the star, the simplest expectation is that the
hottest effective temperature component of the system should present
the bluest colors. 
The evidence here is somewhat equivocal: the Core is blue in 
{\sf k-pahcs}, and adjacent to the bluest part of the {\sf h-k} map, 
but clearly there are more complicated effects going on.
To understand the color maps further, the complexities of scattering 
and reddening in an inhomogeneous environment must be more carefully
modeled.

\begin{figure*}[htb]
\includegraphics[angle=90,width=15cm]{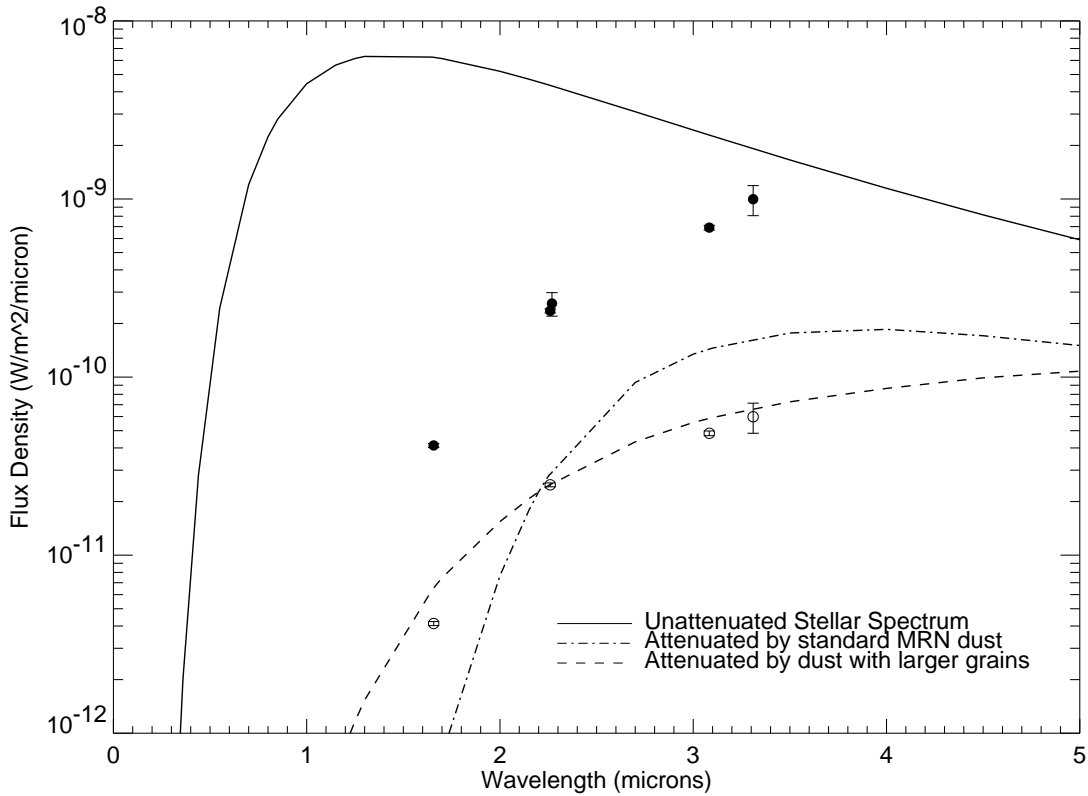}
\caption{ \label{coresed} This figure shows the near-infrared spectral
energy distribution of IRC +10216 as observed by the Keck aperture
masking experiment in 1997 December (filled circles).  The photometry
of the bright compact Core (component ``A'') was isolated from the
surrounding nebula with high resolution techniques, and is also plotted 
(open circles).  
In addition to these data, a 2000K blackbody with uniform disk diameter 
52~mas is shown (solid line).  
This stellar spectrum was reddened by standard MRN dust 
(0.005$\mu$m$<$a$<$0.25$\mu$m, dot-dashed line) and MRN dust with
larger grains (0.005$\mu$m$<$a$<$0.75$\mu$m, dashed line), in both cases
$\tau_{2.2}=5.3$.}
\end{figure*}

Lastly, we subject the hypothesis that component ``A'' is the stellar
photosphere to a final quantitative test.  Assuming the {\it k}~band 52~mas 
Core is the stellar size, and if we adopt a reasonable effective
temperature of 2000~K, the unreddened stellar flux density can be
calculated.  This flux density can be reddened by amorphous carbon
dust \citep{RM1991}, and the resulting SED compared to the
photometry of component ``A'' derived from the Keck aperture masking
frames (calibrated using the contemporaneous interferometric
calibrator and the visibility decomposition from
Figure~\ref{viscurves}).  Figure~\ref{coresed} shows the results of
reddening the stellar spectrum with MRN dust \citep{MRN} using
the standard dust size range (0.005$\mu$m$<$a$<$0.25$\mu$m), as calculated
using DUSTY code \citep{dusty}.   The {\it k}~band flux density of ``A'' is 
well-fitted using $\tau_{2.2} = 5.3$, although the reddened spectrum
is too red to fit the {\it h}~band and L-band Core photometry.  By including somewhat
larger grains (0.005$\mu$m$<$a$<$0.75$\mu$m), all the near-IR photometry
of ``A'' is well-fitted (also, $\tau_{2.2} = 5.3$).   This 
result is not sensitive to the adopted stellar effective temperature.

This simple test does not prove that component ``A'' is the star, however the 
hypothesis does gain credence.  Furthermore, our results that $\tau_{2.2}=5.3$ 
and the finding that the photosphere contributes 10\% of {\it k}~band 
emission is remarkably close to the estimate of \citet{Keady1988} who found
$\tau_{2.2}=5.4$ from SED modeling 
and a photospheric contribution $\sim$12\% confirmed by
dilution of first-overtone CO absorption lines.
Lastly, this line-of-sight dust absorption is consistent (within factor of two)
with the mid-infrared {\em emission} modeled by
\citet{monnier_irc}, based on fits to
the mid-IR spectrum and long-baseline interferometer (ISI) visibility data.

Although these arguments can certainly place the idea that  component ``A'' 
is the star in a favorable light with respect to the known properties of
the system, they cannot be turned around to rule out the alternate hypothesis
that it is a dust clump. 
The observed color temperature, given a number of plausible opacity and 
scattering geometries, could certainly fit with cool circumstellar material. 
Indeed, the finding that the Core is non-circular would seem to imply that
at least some of the light arises from asymmetric circumstellar material,
and/or that there are strong opacity gradients in the line of sight to the star.

In conclusion, although complicated models in which the star is obscured and 
eventually buried in the Northern regions cannot be ruled out, the simple 
alternate hypothesis that Core component ``A'' is highly-reddened, direct 
photospheric light can naturally explain the observed properties of the 
compact Core including: 
a) its angular size, b) its {\it k}~band flux density, c) its color temperature, 
d) its persistence over multiple epochs,  e) the observed dilution of 
near-IR photospheric absorption lines, f) the general level of 
mid-infrared emission, and g) the magnitude of proper motion 
between the Core and North-East arm (``A'' and ``B'').
We encourage detailed radiative transfer modeling constrained by the latest
multi-wavelength imaging to explore the physical structure of this 
fascinating source.



\acknowledgments

Devinder Sivia kindly provided the maximum entropy mapping program ``VLBMEM'', 
which we have used to reconstruct our diffraction limited images. 
This work has been supported by grants from the National Science Foundation 
and the Australian Research Council. The data presented herein were
obtained at the W.M. Keck Observatory, which is operated as a
scientific partnership among the California Institute of Technology,
the University of California and the National Aeronautics and Space
Administration.  The Observatory was made possible by the generous
financial support of the W.M. Keck Foundation.

\clearpage

\end{document}